\documentclass[12pt]{article}%
\usepackage{citesort}
\usepackage{amsmath}
\usepackage{graphicx}%
\usepackage{amsfonts}%
\usepackage{amssymb}
\def\ba{\begin{eqnarray}}
\def\ea{\end{eqnarray}}

\begin{document}

\title{Heavy ion collisions in the used nucleon model}
\author{Adam Bzdak\thanks{Fellow of the Polish Science Foundation (FNP) scholarship
for the year 2006}\\M.Smoluchowski Institute of Physics \\Jagellonian University, Cracow\footnote{Address: Reymonta 4, 30-059 Krakow,
Poland; e-mail: bzdak@th.if.uj.edu.pl}}
\maketitle

\begin{abstract}
It is shown that recently proposed by R.J. Glauber the used nucleon model
combined with the assumption that the nucleon consists of two constituents (a
quark and a diquark) describes well the PHOBOS data on particle production at
midrapidity.



\end{abstract}


\vskip 0.8cm

\textbf{1.} In this note we investigate the consequences of the used nucleon
model \cite{unm} in the context of the PHOBOS data on particle production at
midrapidity \cite{AuAu-130,AuAu-19-200,AuAu-62}. This model was formulated as
a simple generalization of the wounded nucleon model \cite{wnm} which fails to
describe correctly the observed particle multiplicities.

Contrary to the wounded nucleon model, we assume that the number of particles
produced from one wounded nucleon does depend on the number of inelastic
collisions this nucleon underwent. To formulate this model in more detail let
us, for a while, consider nucleon-nucleus (mass number $A$) collision. For the
incident nucleon contribution we have: the first inelastic collision produces
$n$ particles, where $2n\equiv(dn/d\eta)|_{|\eta|<1}$ is the average
multiplicity at midrapidity in a single proton-proton collision. Following
\cite{unm} we assume that the second collision produces a fraction $\mu$ of
$n$, the third one $\mu^{2}$, and so on. The number of particles produced
(from the incident nucleon) after $k$ collisions is:%
\begin{equation}
n_{k}\equiv n\left(  1+\mu+...+\mu^{k-1}\right)  =n\frac{1-\mu^{k}}{1-\mu}.
\label{szereg}%
\end{equation}

From (\ref{szereg}) it is seen that the used nucleon model gives a natural way
of interpolating between the two limits: for $\mu=0$ we arrive at the wounded
nucleon model \cite{wnm} i.e. there is no difference whether a nucleon is hit
once or several times, $n_{k}=n$; for $\mu=1$ the incident nucleon
contribution is proportional to the number of collisions, $n_{k}=nk$.

The probability $P_{k}$ that the incident nucleon underwent exactly $k$
inelastic collisions is given by a standard formula:%
\begin{equation}
P_{k}(b)=\left(
\genfrac{}{}{0pt}{}{A}{k}%
\right)  \left[  \sigma_{NN}T_{NN}(b)\right]  ^{k}\left[  1-\sigma_{NN}%
T_{NN}(b)\right]  ^{A-k}, \label{p_k}%
\end{equation}
where $T_{NN}$ is given by:%
\begin{equation}
T_{NN}(b)=\frac{1}{\sigma_{NN}}\int\sigma_{NN}\left(  s\right)  T(b-s)d^{2}s.
\label{T_NN}%
\end{equation}
Here $\sigma_{NN}$ and $\sigma_{NN}(s)$ are the total and differential
inelastic proton-proton cross-sections, respectively. $T(s)$ is the thickness
function (normalized to unity).

The average number of particles produced from the incident nucleon is given
by:%
\begin{equation}
\sum\nolimits_{k}P_{k}n_{k}=\frac{n}{1-\mu}\left\{  1-\left[  1-\left(
1-\mu\right)  \sigma_{NN}T_{NN}(b)\right]  ^{A}\right\}  . \label{<n>}%
\end{equation}

Consequently, the average multiplicity at midrapidity $N\equiv(dN/d\eta
)|_{|\eta|<1}$ for the symmetric nucleus-nucleus collision at a given impact
parameter $b$ one obtain%
\begin{equation}
N(b)=\frac{2A}{\sigma_{AA}(b)}\frac{n}{1-\mu}\int T(b-s)\left\{  1-\left[
1-\left(  1-\mu\right)  \sigma_{NN}T_{NN}\left(  s\right)  \right]
^{A}\right\}  d^{2}s, \label{M_Au}%
\end{equation}
where $\sigma_{AA}(b)$ is the inelastic differential nucleus-nucleus
cross-section.\footnote{In case of Au-Au collisions, using the optical
approximation, we have verified that $\sigma_{AA}(b)=1$, for $b<14$ fm.}

\textbf{2.} The PHOBOS data are presented versus the number of wounded
nucleons in both colliding nuclei $W$, given by \cite{wnm}%
\begin{equation}
W(b)=\frac{2A}{\sigma_{AA}(b)}\int T(b-s)\left\{  1-[1-\sigma_{NN}%
T_{NN}(s)]^{A}\right\}  d^{2}s, \label{W}%
\end{equation}

Here and in (\ref{M_Au}) for the nuclear density we use the standard
Woods-Saxon formula with the nuclear radius $R_{Au}=6.37$ fm and $d=0.54$ fm.
We assume the differential proton-proton cross-section $\sigma_{NN}\left(
s\right)  $ to be in a Gaussian form:\footnote{We performed the full
calculation also in case of point-like interaction approximation, $\sigma
_{NN}(s)=\sigma_{NN}\delta^{2}(s)$. We have observed that centrality
dependence is slightly worse, indicating that the more complete treatment is
indeed needed.}%
\begin{equation}
\sigma_{NN}\left(  s\right)  =\gamma e^{-s^{2}/\varkappa^{2}},
\label{sigma_NN}%
\end{equation}
where $\varkappa^{2}=\sigma_{NN}/(\pi\gamma)$ and $\gamma=0.92$
\cite{p(0),ab-ab}.

We have observed that the used nucleon model gives a good description of the
RHIC Au-Au data with $\mu\sim0.41$.

\begin{figure}[h]
\begin{center}
\includegraphics[scale=0.97]{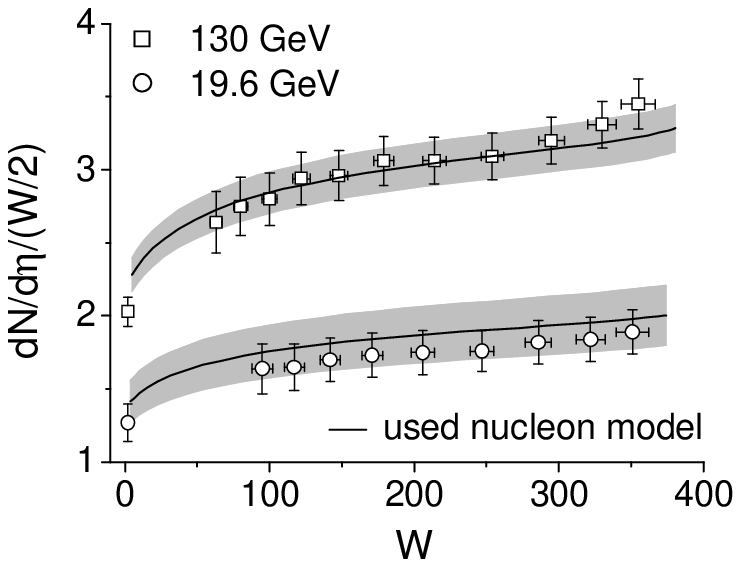}  \hspace{0.2cm}
\includegraphics[scale=0.97]{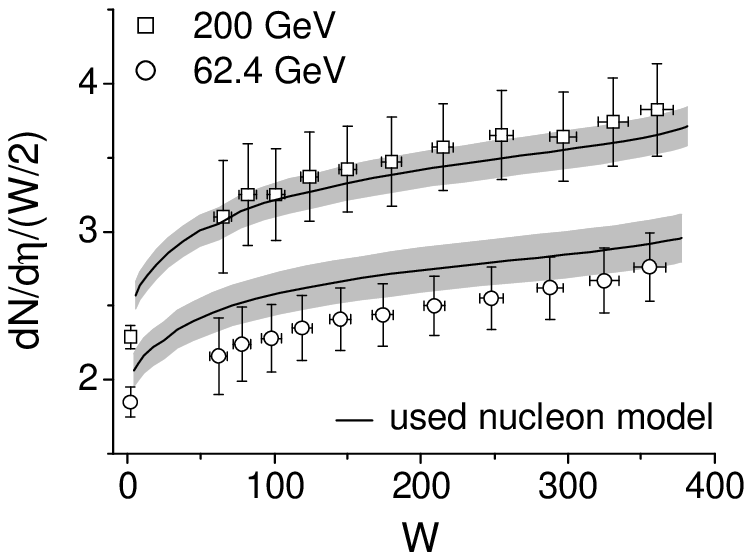}
\end{center}
\caption{The predictions of the used nucleon model (for $b\leq14$,
corresponding to $W\gtrsim5$) with $\mu=0.41$ compared with the PHOBOS\ data
\cite{AuAu-130,AuAu-19-200,AuAu-62}. The shaded areas reflect the inaccuracy
in the inelastic pp data.}%
\label{AA_fig}%
\end{figure}

The comparison of the used nucleon model ($\mu=0.41$) with the PHOBOS data
\cite{AuAu-130,AuAu-19-200,AuAu-62} on the average multiplicity at midrapidity
per one wounded nucleon as a function of $W$ is shown in Fig. \ref{AA_fig}.
Also shown are inelastic pp data (points at $W=2$), as quoted in
\cite{AuAu-19-200,AuAu-62}.

The inelastic proton-proton cross-sections needed for this calculation were
taken as $\sigma_{NN}=32$ mb, $36$ mb, $41$ mb and $42$ mb at the RHIC
energies $\sqrt{s}=19.6,$ $62.4,$ $130$ and $200$ GeV, respectively.

\textbf{3.} To interpret the parameter $\mu$ let us consider two simple
production scenarios.

(i) Wounded quark model \cite{wqm}. It is assumed that the nucleon consists of
three constituent quarks and particle production from these quarks is
independent on the number of interactions they underwent. Furthermore, we
assume that the number of produced particles after $k$ collisions is
proportional to the number of wounded quarks, that is:%
\begin{equation}
\frac{n_{k}}{n}=\frac{w_{k}^{(q)}}{w^{(q)}}, \label{prop}%
\end{equation}
where $w^{(q)}$ is the average number of wounded quarks per nucleon in a
single inelastic proton-proton collision, $w_{k}^{(q)}$ is the average number
of wounded quarks in the nucleon which underwent $k$ inelastic collisions. The
latter may be calculated by a straightforward counting of probabilities:
\begin{equation}
w_{k}^{(q)}=3\left[  p_{q}+p_{q}(1-p_{q})+...+p_{q}(1-p_{q})^{k-1}\right]  .
\label{w-k-q}%
\end{equation}
Here $p_{q}$ is the probability for a quark to interact in a single
proton-proton collision (note that $p_{q}=w^{(q)}/3$). Taking (\ref{prop}) and
(\ref{w-k-q}) into account we obtain:%
\begin{equation}
n_{k}=n\left[  1+(1-p_{q})+...+(1-p_{q})^{k-1}\right]  =n\frac{1-(1-p_{q}%
)^{k}}{p_{q}}. \label{mi-3q}%
\end{equation}
Comparing (\ref{mi-3q}) with (\ref{szereg}) we find that $\mu=1-p_{q}$.\ 

It turns out that at the RHIC energies approximately $w^{(q)}\approx1.2$
\cite{li-zr-3q}, giving $\mu\approx0.6$. We checked that this value leads to
significantly larger multiplicities than actually observed \cite{bw}%
.\footnote{In order to obtain the value of $\mu\simeq0.41$ we would have to
assume that $w^{(q)}\approx1.75$. This number, however is difficult to justify.}

(ii) Wounded quark-diquark model \cite{ab-ab}. We assume that the nucleon is
composed of two constituents (a quark and a diquark). As above we assume that
particle production from these constituents is independent on the number of
interactions they underwent, and both constituents produce the same number of
particles. Performing analogous calculations as above we obtain:%
\begin{equation}
n_{k}=n\frac{1-(1-p_{q})^{k}}{p_{q}+p_{d}}+n\frac{1-(1-p_{d})^{k}}{p_{q}%
+p_{d}}, \label{n-k-qd}%
\end{equation}
were $p_{q}$ and $p_{d}$ are the probabilities for a quark and a diquark to
interact in a single proton-proton collision, respectively.

As it is not clear what the diquark is, however, we may consider two different possibilities:

(a) $p_{d}=p_{q}$, it corresponds to the situation where both constituents are
\textit{similar}.\footnote{It may mean that both constituents are of the same
size.} In this case (\ref{n-k-qd}) reduces to (\ref{szereg}) with $\mu
=1-p_{q}$. Assuming the average number of wounded constituents in a single
proton-proton collision $w^{(q+d)}=$ $p_{q}+p_{d}$ to be $1.18\div1.19$
\cite{ab-ab} we obtain the proper value $\mu\approx0.41$.

(b) $p_{d}=2p_{q}$, where diquark is rather large, comparable to the size of
the proton (this relation was obtained in \cite{ab-ab}). Now it is not
possible to use formula (\ref{M_Au}), however, performing analogous
calculations leading to (\ref{M_Au}) we obtain:%
\begin{align}
N(b)  &  =\frac{2A}{\sigma_{AA}(b)}\frac{n}{w^{(q+d)}}\int T(b-s)\left\{
1-\left[  1-p_{q}\sigma_{NN}T_{NN}\left(  s\right)  \right]  ^{A}+\right.
\label{N}\\
&  +\left.  1-\left[  1-p_{d}\sigma_{NN}T_{NN}\left(  s\right)  \right]
^{A}\right\}  d^{2}s.\nonumber
\end{align}

We have checked that this formula with $p_{d}=2p_{q}$ and $w^{(q+d)}=$ $1.185$
\cite{ab-ab} gives the results which practically do not differ (less than $5$
$\%$) than those presented in Fig. \ref{AA_fig}. Thus, we may conclude that
the model is almost independent on the details of the diquark\footnote{We also
checked for other choices between $p_{d}=p_{q}$ and $p_{d}=2p_{q}$.}, and the
only thing that really matters is the number of constituents.

\textbf{4.} In conclusion, we have shown that the used nucleon model combined
with the assumption that the nucleon consists of two constituents (a quark and
a diquark) naturally describes the PHOBOS data on particle production at midrapidity.

\vskip 0.3cm

\textbf{Acknowledgements}

I would like to thank prof. Andrzej Bialas for discussions.


\begin{thebibliography}{9}                                                                                                %

\bibitem {unm}R. J. Glauber, Nucl. Phys. A774 (2006) 3.

\bibitem {AuAu-130}B. B. Back et al., Phys. Rev. C65 (2002) 061901.

\bibitem {AuAu-19-200}B. B. Back et al., Phys. Rev. C70 (2004) 021902.

\bibitem {AuAu-62}B. B. Back et al., Phys. Rev. C74 (2006) 021901.

\bibitem {wnm}A. Bialas, M. Bleszynski, W. Czyz, Nucl. Phys. B111 (1976) 461.

\bibitem {p(0)}U. Amaldi, K. R. Schubert, Nucl. Phys. B166 (1980) 301.

\bibitem {ab-ab}A. Bialas, A. Bzdak, nucl-th/0611021.

\bibitem {wqm}A. Bialas, W. Czyz, W. Furmanski, Acta Phys. Pol. B8 (1977) 585;
A. Bialas, W. Czyz, L. Lesniak, Phys. Rev. D25 (1982) 2328.

\bibitem {li-zr-3q}V. V. Anisovich, M. N. Kobrinsky, J. Nyiri, Yu. M.
Shabelsky, Quark Model and High Energy Collisions, World Scientific,
Singapore, 1985.

\bibitem {bw}B. Wosiek, private communication; H. Bialkowska, nucl-ex/0609006.
\end{thebibliography}
\end{document}